\documentclass{elsart}

\usepackage{graphicx}
\graphicspath{{./images/}}
\usepackage{cite}
\usepackage{amsmath}
\usepackage{amssymb}
\usepackage[latin1]{inputenc}

\begin{document}

\begin{frontmatter}

\title{Magnetism of 3d transition metal atom on Au(110)-(1$\times2$) and Au(111) surfaces}
\author[a]{Wei Fan},
\author[b,a]{Xin-Gao Gong},
\address[a]{ Key Laboratory of Materials Physics,
 Institute of Solid State Physics, Chinese Academy of Sciences, 230031-Hefei, P. R. China}
\address[b]{Department of Physics, Fudan University, 200433-Shanghai, P. R. China}

\date{\today}

\begin{abstract}
  We calculate the magnetism of 3d transition-metal atoms on Au(110)-(1$\times$2) and Au(111) surfaces
  based on the Density Functional Theory.  Our results show that the surface relaxation
  enhances the orbital moments of left-end elements (Ti,V) and quenches the orbital moments of right-end
  elements (Fe,Co,Ni) on the Au(110)-(1$\times$2) surface. The middle elements (Cr,Mn) of the group have large
  spin moments on the two surfaces because of the strong electronic exchanged and correlated interactions.
\end{abstract}


\end{frontmatter}

\section{Introduction}
\label{sec:intro}

 The magnetism of nano-structures on a surface is a current interesting topic in material
 science and surface science due to their potential applications to the high-density magnetic
 recording and the memory devices. The magnetism of a single magnetic atom on a surface is
 important to understand the magnetism of materials with more complex structures. The reduced
 dimensionality and coordination number for surface enhance the magnetism of
 surface and absorbed atoms. The magnetism of absorbed atoms have been studied in
 experiments~\cite{6Gross1,6Gross2,Hanf1,6Riegel1,Ortega1}, theoreticals model and
 calculations of (DFT) density functional theory~\cite{Stepanyuk1,6Wildberger1,6Stepanyuk2,6Cabria1,6Hjortstam1}.

 The magnetism of materials has close relation with material structure and vice versa, such as large
 magnetic moment stabilizes structures of clusters doped with magnetic impurities~\cite{6Gong1}.
 Both experiments and theories have reported that the atoms absorbed on surface have large orbital moments
 compared with the atoms in the bulk~\cite{6Cabria1,6Hjortstam1}. Besides the reduced coordination number,
 the broken global symmetry (Space group) can also enhance orbital moments of the absorbed atoms, such as
 the broken from three-dimensional translation symmetries in the bulk to the two-dimensional translation
 symmetries on the surface makes the orbital quenching uncompleted. The broken of local symmetries (Point group)
 generally quenches the orbital moments efficiently because the lost of local symmetries can efficiently lift
 the degeneracy of the ground state. The orbital polarization, spin-orbit coupling and electronic correlated
 interaction also have significant contributions to orbital moments. Some of DFT calculations with the orbital
 polarization term have obtained correct orbital moments of the atoms absorbed on
 surface~\cite{Stepanyuk1,6Wildberger1,6Stepanyuk2,6Cabria1,6Hjortstam1}. However the orbital polarization generally
 overestimates the magnetic anisotropy energy (MAE) of magnetic materials~\cite{6Hjortstam1,6Trygg1}.

 The surface deformation induced by absorbed atoms generally changes the symmetries and strength of
 crystal field. The orbital moment will have significant changes because orbital
 moment is close related to the symmetries and strength of the crystal field.
 In this work we calculate the magnetism of a single absorbed 3d atom on Au(110)-(1$\times$2) and Au(111)
 surfaces. The Au(110)-(1$\times$2) surface reconstructs from the Au(110) surface by missing Au rows every
 others and has one-dimensional troughs along the closed packed [1$\bar{1}$0] direction.
 Our results show the orbital moments of the left-end elements Ti and V increase if they relax together
 with surface and decrease for the left-end elements such as Co and Ni. The spin moments of the two-side
 elements of the 3d group such as Ti, V, Co and Ni have smaller values than those of the middle elements
 such as Cr and Mn. The absorbed 3d atoms in the middle of this group still keep the large spin moments
 and close to their values of individual atoms due to the strong electronic exchanged and correlated
 interactions.

\section{Computation Methods}
\label{sec:method}

\begin{figure}
 \centering
 \includegraphics[width=16cm]{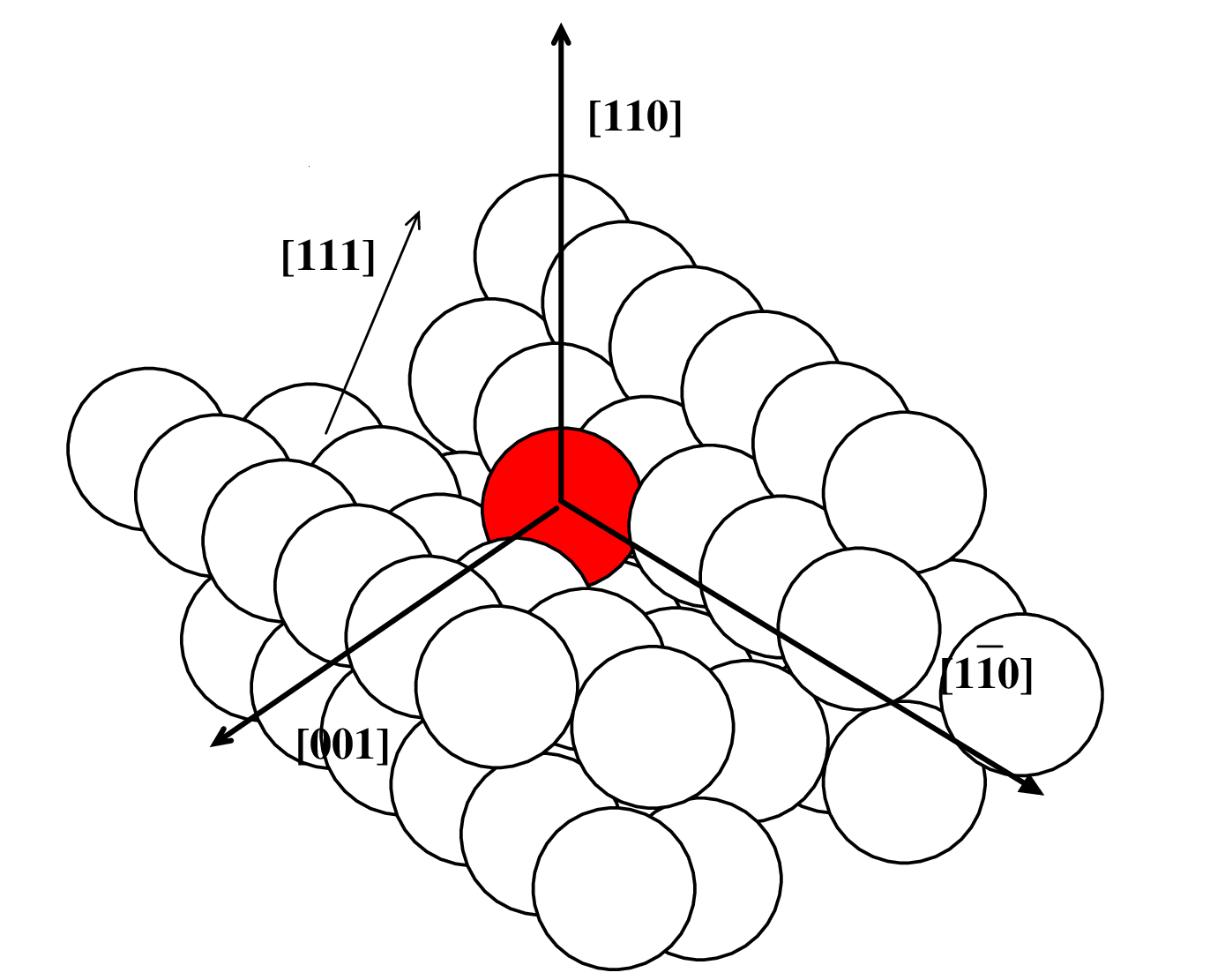}
 \caption{  The schematic diagram of a 3d transition metal atom absorbed
 on the Au(110)-(1$\times$2) surface. The trough is along the [1$\bar{1}$0] direction.
 The normal of the facet of the trough is the [111] direction. The adatom absorbs on
 an hollow site in the trough. The white spheres are the Au atoms of the surface, the
 gray sphere the absorbed adatom.}
 \label{fig1}
 \end{figure}

 We calculate the magnetism of the 3d transition metal atoms from Ti to Cu which are absorbed on
 Au(110)-(1$\times$2) and Au(111) surfaces based on the Density Functional Theory~\cite{Hohenberg1,Kohn1}
 using the Methods of Projection of Augmentation Wave (PAW)~\cite{Blochl1} with the plane-wave
 base-set and the Perdew-Burke-Ernzerhof GGA exchange-correlate potential~\cite{Perdew1}. The program is
 the VASP code written by the computational materials science group at Vienna
 university~\cite{Kresse1,Kresse2,Kresse3,Hobbs1}. The initial structures are constructed by placing the 3d
 atom on the hollow sites of the Au(111) and Au(110)-(1$\times$2) surface in the trough (Fig.~\ref{fig1}).
 The initial surface constructs with 28 Au atoms with 4 atomic layers. The atoms in the layer of the
 slab bottom are unrelaxed. At the first step we optimized the lattice constant 4.079$\AA$ of the slab
 by having changed the lattice constant and kept the slab unchanging. The total-energy
 minimum applied the RMM-DIIS algorithm~\cite{Kresse1} with plane-wave energy cutoff 269.561eV and
 the 6x6x6 Monkhorst-Pack K points. The size of the super-cell is
 (11.523$\AA$$\times$8.148$\AA$$\times$15.52$\AA$).
 The initial structures are relaxed using the conjugate-gradient Methods with $\Gamma$
 points used as the Brillouin-zone sampling in the corresponding calculations of the
 electronic structure. Based on the relaxed structures, we calculate the electronic structures
 of Ni chains using the RMM-DIIS algorithm~\cite{Kresse1} with the (6$\times$6$\times$1) Monkhorst-Pack
 grids sampling the Brillouin zone. If the changes of the total energies are smaller than 0.0001eV between
 two electronic self-consistent (SC) steps the SC-loops break, and 0.001eV between two ion-steps
 the programs stop. The Wigner-Seitz radiuses 1.323$\AA$ is set for Ti, V, Cr, Mn, 1.302$\AA$ for Fe and Co
 1.286$\AA$ for Ni, 1.312$\AA$ for Cu to calculate the magnetic moments. The energy cutoffs of plane waves are
 set to 229.9eV for Ti, V and Cr, 269.9eV for Mn, 267.9eV for Fe, 268.0eV for Co, 269.6eV for Ni, 273.2eV for
 Cu atoms. We set the Methfessel-Paxton smearing width equal to 0.20eV to accelerate the speed of convergence.
 We choose the [110] direction as the quantization axis.

 We emphasize the influence of surface deformation on the orbital moments of the absorbed atoms.
 The orbital polarization interaction omits in this work but still include the spin-orbit coupling
 interaction. We obtain smaller orbital moments compared with pervious DFT calculation including
 the orbital polarization interaction~\cite{6Nonas1}.

\section{Results and Discussion}
\label{sec:result}

\begin{figure}
 \centering
 \rotatebox{180}{\includegraphics[width=16cm]{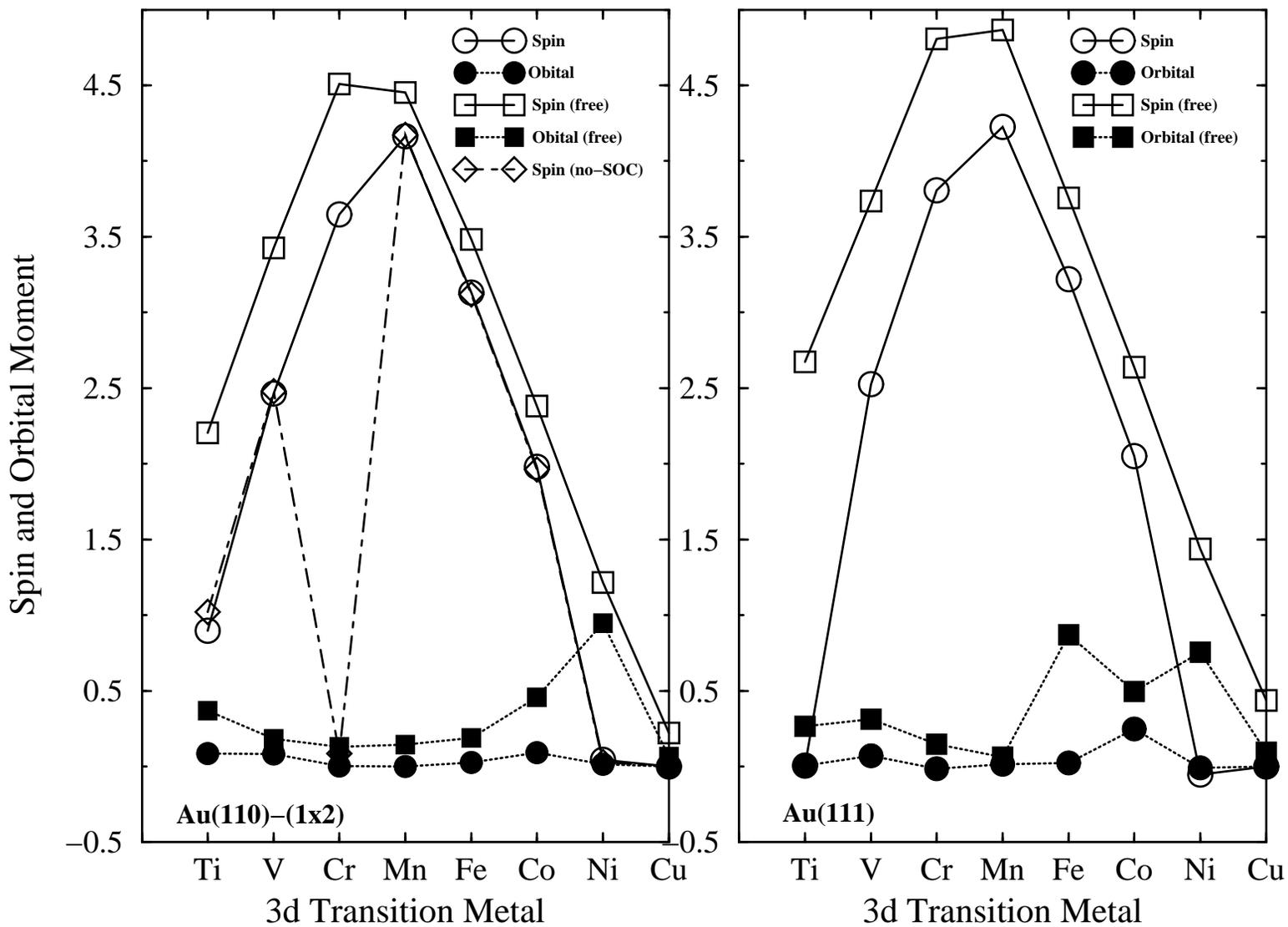}}
 \caption{The spin (solid lines) and orbital (dot
 lines) moments of the absorbed 3d atoms and the free standing 3d atoms.
 The left panel shows the results for Au(110)-(1$\times$2) surface, the right
 panel for the Au(111) surface. The dot-dash line shows the collinear result
 for Au(110)-(1$\times$2) surface.}
 \label{fig2}
 \end{figure}

 At first we calculate the magnetic moments of the absorbed 3d atoms on the Au(110)-(1$\times$2) and
 Au(111) surfaces with the spin-orbit coupling interactions and non-collinear calculations\cite{6Fan1}.
 We also calculate the magnetism of the free standing 3d atoms by removing the surface atoms and keeping the
 other parameters unchanged. From Fig.~\ref{fig2}, we can see that the spin moments reach
 the maximum in the middle of the group. The large spin moment of the absorbed Mn atom
 is about 4.163$\mu_{_{B}}$ on the Au(110)-(1$\times$2) surface. We can see from the left panel of
 the figure that the spin moment of the absorbed Cr atom is the second largest when including the
 spin-orbit coupling interactions and non-collinear calculations, and almost zero without including them.

 Our results show that the spin moments of the absorbed atoms are smaller than those of the free
 standing atoms (Fig.~\ref{fig2}) due to the hybridizing interactions. The magnetic moments of the
 absorbed atoms near the two ends of 3d elements group such as Ti,V,Cr,Ni decrease
 the larger values compared with the middle atoms of the group such as Mn,Fe,Co. Thus our results
 show that the hybridization weakens the efficient of the Hund rule. The localized 3d states of the
 absorbed 3d atoms will be delocalized when they are embedded in the environments of the extended
 surface states. The weakened coulomb interactions between electrons in the delocalized states will
 produce more double-occupations and decrease spin moments of the absorbed 3d atoms. Our results
 also show that the absorbed Cr and Mn atoms are more like free standing atoms than the other 3d atoms
 because of stronger exchanged and correlated interactions. The decreases of their spin moments are
 smaller than those of the other 3d atoms due to the hybridization with surface.
 We can find that the spin moment of the absorbed Ti atom decreases almost to zero when it absorbs
 on the Au(111) surface. The changes for the spin moments on the Au(111) surface
 are similar to those on the Au(110)-(1$\times$2) surface across the group.

 Both the electronic correlations and the crystal field have significant influence on the orbital moments of
 the absorbed 3d atoms. If the electronic correlation is stronger than the crystal fields, the orbital moment
 is large, otherwise small. The absorbed Ti,V,Co atoms have visible orbital moments
 0.086$\mu_{_{B}}$,0.082$\mu_{_{B}}$ and 0.092$\mu_{_{B}}$ on the Au(110)-(1$\times$2) surface, and
 0.071$\mu_{_{B}}$ and 0.247$\mu_{_{B}}$ on the Au(111) surface respectively. Cabria, et.al obtained larger
 orbital moment (about 0.5$\mu_{_{B}}$) of Fe and Co atoms absorbed on Au(001) surface using the spin
 polarization relativistic KKR methods including the  orbital polarization term~\cite{6Nonas1}. The usual
 exchange-correlate potentials (LDA or GGA) underestimate the electronic correlations such as the Coulomb
 correlation and the orbital polarization, thus the obtained orbital moments in this work are generally small
 compared with experimental values~\cite{6Brooks1,6Solovyev1,6Eriksson1,6Nonas1} although our
 work includes the spin-orbital coupling interaction. The small values of orbital moments of the absorbed Cr
 atoms in our calculations are not related to the crystal field but to its electronic structure. In the
 individual Cr atom, five 3d-electrons half-fill the 3d states, the total orbital moment is very small.
 The large spin moment, small orbital moment and the weak interaction with the surface all imply that the
 absorbed Cr atom is just like an individual atom.

 Generally, the material structures are close related to the material magnetism.
 The smaller atomic nearest neighbor distance leads to the larger exchanged integral $J$.
 The strong magnetic correlation of different atoms enhances the magnetism of the materials.
 The distances of surface atoms are generally smaller than those in bulk. Thus the magnetism of
 surface are generally stronger than that in bulk. The supported atoms on the surface
 are in the environment similar to the surface atoms. Based on the same logic, the supported atoms
 on the surface have possibly strong magnetism.

\begin{figure}
 \centering
 \includegraphics[width=15cm]{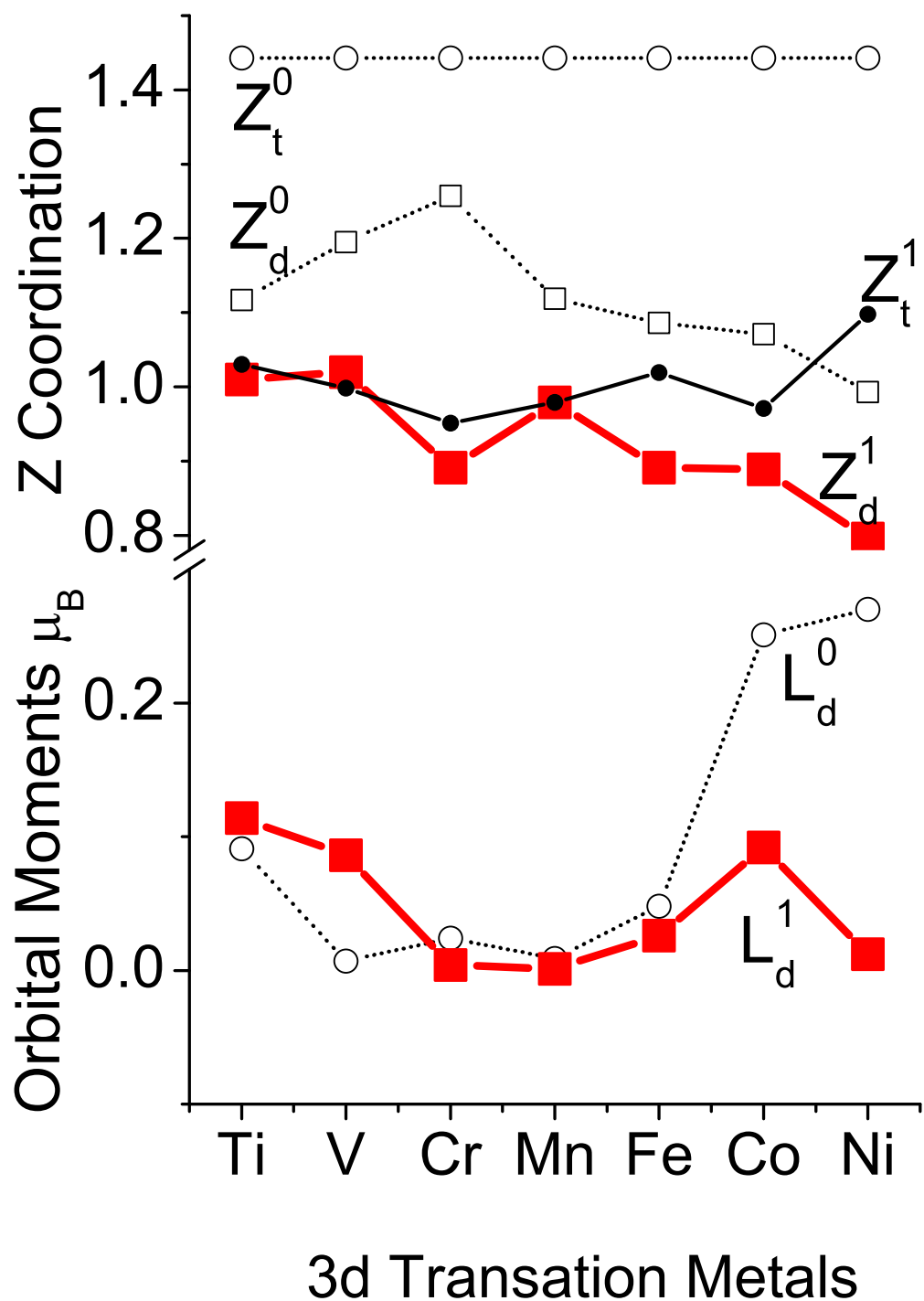}
 \caption{The effects of relaxation on the orbital moments of
 the absorbed 3d atoms. L$^{0}_{d}$ and L$^{1}_{d}$ are the orbital moments before and after
 relaxations illustrated by the short dots line with open circles and the solid line with filled
 squares. Z$^{0}_{d}$ and Z$^{0}_{t}$ (superscript 0) are the average Z coordinates before
 relaxations, and Z$^{1}_{d}$ and Z$^{1}_{t}$ (superscript 1) after relaxations. The subscript
 'd' represents the absorbed atoms and 't' the top-row atoms of the reconstructed Au(110) surface.
 The Z coordinates measure from the second surface layer.}
 \label{fig3}
 \end{figure}

 The lack of the orbital polarization isn't the obstacle to the study of the structure influence on
 the orbital moments.
 In order to clearly illustrate the effects of the crystal deformations (or the cubic distortions),
 we compare the results on the deformed surface with that on the prefect Au(110)-(1$\times$2) surface.
 The absorbed 3d atoms on the prefect surface still modify their positions to reach their stable positions,
 although the surface atoms are fixed. We find from Fig.~\ref{fig3} that the surface relaxations generally quench the orbital
 moments of the absorbed 3d atoms except for the absorbed V and Ti atoms. The changes of the orbital moments
 in response to the relaxations are closely related to the changes of the depth of the 3d
 atoms embedded in the trough of the reconstructed Au(110) surface. The absorbed atoms are deeper in the through,
 their orbital moments are quenched to smaller values due to the stronger crystal filed.
 The absorbed V atom rises above the top row of the trough after the relaxations and its orbital moment
 enhances, opposes to the orbital quenching for the other 3d atoms except for the absorbed
 Ti atom. This is due to the weaker crystal field above the surface than that in the trough.
 The absorbed Ti atom has almost the same height as the top row after the relaxations and its orbital
 moment slightly increases. Thus our results indicate that the surface relaxation decreases the orbital
 moments of the absorbed 3d atoms with the excess half-filled 3d states and increases the orbital
 moments of the absorbed atoms with less half-filled 3d states on Au(110)-(1$\times$2) surface.

\begin{figure}
 \centering
 \includegraphics[width=16cm]{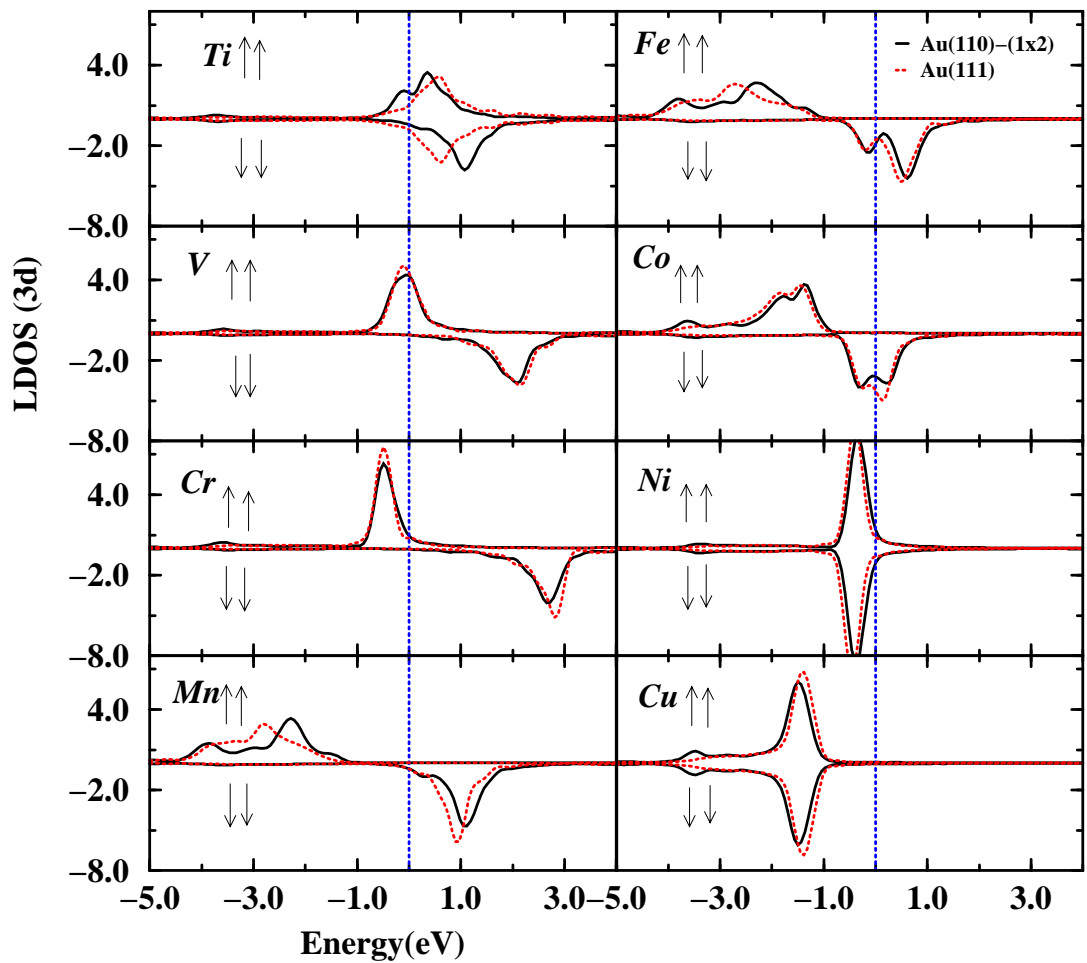}
 \caption{The 3d densities of state (DOS) for the
 3d atoms absorbed on Au(110)-(1$\times$2) (solid lines) and Au(111) (dot lines )
 surface.}
 \label{fig4}

 \end{figure}

 The 3d density of states of the absorbed 3d atoms on the two surfaces are shown in Fig.~\ref{fig4}.
 From this figure, the large spans between the spin-up and spin-down peaks imply the large
 electronic correlations for the Mn, and Cr, and Fe atoms. This is why the magnetism of the
 absorbed Cr, Mn, Fe atoms are similar to the magnetism of their individual atoms. The 3d DOS of
 the absorbed Fe has more contributions to the total DOS at Fermi energies. The electrons near Fermi
 energy form a glue which sticks atoms in bulk together~\cite{2Daw2,2Ercolessi1} and also dominates
 the electron transports of materials. The absorbed Ti, V, Fe and Co atoms with large 3d DOS at the Fermi
 energies all have strong interactions based on the calculations of the interaction energies. On the contrary,
 the absorbed Mn, Cr, Cu atoms with small 3d DOS at the Fermi energies have weak interactions
 with the two surfaces~\cite{6Fan1}. Our results
 also show that the 3d DOS at Fermi energies are also closely related the orbital moments of the
 absorbed 3d atoms. The absorbed Ti, V, Co atoms on the Au(110)-(1$\times$2) surface and V, Co atoms
 on the Au(111) surface with large 3d DOS at the Fermi energies have large orbital moments. We get
 the same arguments on the magnetism of Ni$_{n}$ (n=1-5) chains in our parallel research~\cite{6Fan1}.
 It is valuable to note that we can not infer the large orbital moment for the absorbed Fe atom
 based on our results. It can be rescued by the introductions of the orbital-polarization
 term~\cite{6Brooks1} or the correction of the on-site coulomb energy U.

\begin{figure}
 \centering
 \includegraphics[width=12.0cm]{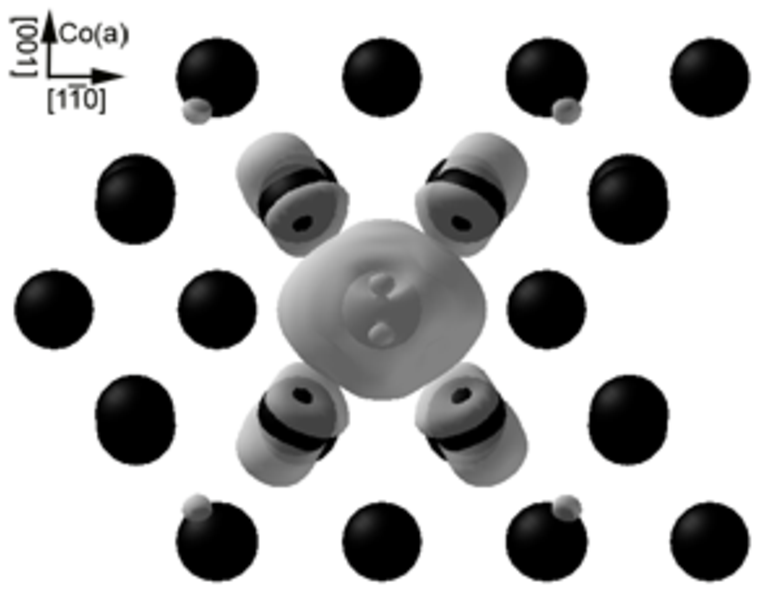}
 \includegraphics[width=12.0cm]{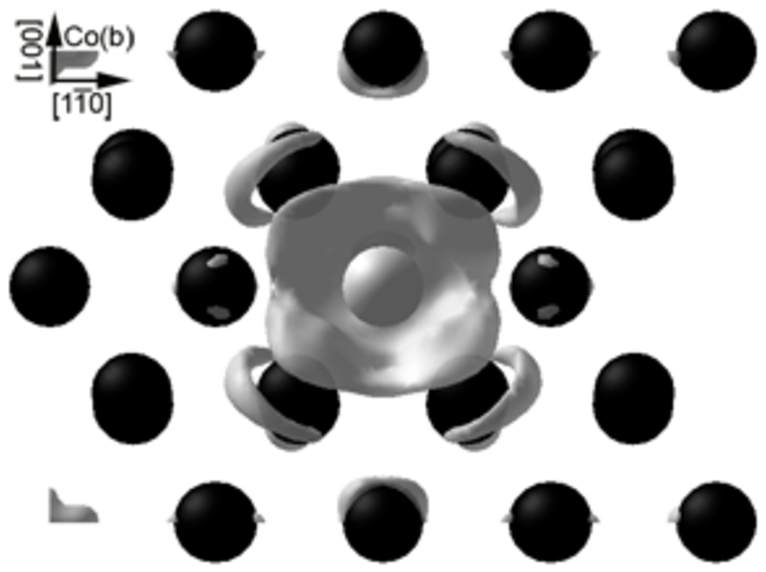}
 \caption{ The magnetic density of the absorbed Co atom which is
 the components along the quanta axis [110] and perpendicular to the surface. The magnetic density is defined
 as $\vec{m}(r)=\Sigma_{\alpha\beta}n^{\alpha\beta}\cdot\vec{\sigma}_{\alpha\beta}$, $\vec{\sigma}$ is the Pauli
 Matrix. The figure shows the component along the quantization axis. The contour values
 are 0.002/$\AA^{3}$ for the left panel and -0.002/$\AA^{3}$ for the right panel}
 \label{fig5}
 \end{figure}

 The STM~\cite{6Jamneala1}experiments of 3d atoms absorbing on Au(111) surface show that the elements
 (Ti, Co, Ni) near the two ends of the group have obviously Kondo resonances and for the middle elements
 not. The intense of STM spectrums at zero bias voltage is proportional to the local density of
 state (LDOS) at the Fermi energy. The STM experiments indicate the large LDOS near the Fermi energy
 for the absorbed Ti, Co and Ni atoms. The density functional theory with the LDA or GGA
 exchange-correlate potential generally doesn't completely include the Kondo correlations. The Kondo
 resonance and Kondo effects origin from the screening of the conducting electrons on the
 local spin~\cite{6Kondo1,6Hewson1}. The ground state is the Kondo single state with total zero
 spin~\cite{6Wilson1}. Thus, the screening effect is an important aspect for the Kondo resonance.
 Although the calculations of DFT can not show the completely screening to the local spin, we still hope
 to find the partial screening. Fig.~\ref{fig5} shows the partial screening effect to the local spin for
 the absorbed Co atom.

\section{Conclusion}
\label{sec:conclusion}

  In conclusion, we calculate the magnetism of 3d transition metals on Au(110)-(1$\times$2)
  surface and Au(111) surface based on the Density Functional Theory with Project Argument
  Wave Methods and plane-wave base-sets. Our results show that the spin moments of the
  two-side elements of the 3d group such as Ti, V, Co and Ni have smaller values than
  those of the middle elements such as Cr and Mn. The absorbed 3d atoms in the middle of this
  group still keep the large spin moment similar to their individual atoms due to the strong electronic
  correlation. The surface relaxations generally deform the surface and modify the positions of the
  absorbed atoms. If the relaxations lift the absorbed atom out of the trough, the orbital moments
  enhance due to weaker lattice fields. If the relaxations make the absorbed atoms deeper embedded
  in the trough the orbital moments quench to smaller values due to the stronger lattice fields .

\section*{Acknowledgements}

 W.Fan thanks Professor Q.Q.Zheng, Z.Zeng and Dr. J.L.Wang for useful discussion.
 W.Fan is supported by  Center for computational science, Hefei Institutes of Physical Science,
 Chinese Academy of Sciences; Nature Science Foundation of China under Grant Nos 10374091 and 90103038,
 Knowledge Innovation Program of Chinese Academy of Sciences under KJCX2-SW-W11. The jobs run on the SGI-3900
 parallel computer at Center for computational science, Hefei Institutes of Physical Science, Chinese Academy
 of Sciences. X.G.Gong is supported by Nature Science Foundation of China, the special funds for major state
 basic research and GAS projects.

\newpage



\begin{thebibliography}{99}
  \expandafter\ifx\csname url\endcsname\relax
  \def\url#1{\texttt{#1}}\fi
  \expandafter\ifx\csname
  urlprefix\endcsname\relax\def\urlprefix{URL }\fi

 \bibitem{6Gross1} K.D.Gross,D.Riegel,and R.Zeller,
  Phys. Rev. Lett. {\bf 63} (1989) 1176.
 \bibitem{6Gross2} K.D.Gross,D.Riegel,and R.Zeller,
  Phys. Rev. Lett. {\bf 65} (1990) 3044.
 \bibitem{Hanf1} M.C.Hanf, C.Pirri, J.C.Peruchetti, D.Bolmont, and G. Gewinner,
  Phys. Rev. B{\bf 36} (1987) 4487.
 \bibitem{6Riegel1} D.Riegel,L.B\"uermann,K.D.Gross,M.Luszik-Bhadra, and S.N.Mishra,
  Phys. Rev. Lett. {\bf 62} (1989) 316.
 \bibitem{Ortega1} J.E.Ortega and F.J.Himpsel
  Phys. Rev. B{\bf 47} (1993) 16441.
 \bibitem{Stepanyuk1} V.S.Stepanyuk, A.N.Baranov, W.Hergert and P.Bruno,
  Phys. Rev. B{\bf 68} (2003) 205422.
 \bibitem{6Wildberger1} K. Wildberger, V.S.Stepanyuk, P. Lang, R. Zeller, and P. H. Dederichs,
  Phys. Rev. Lett. {\bf 75} (1995) 509.
 \bibitem{6Stepanyuk2} V.S.Stepanyuk, W.Hergert, P. Rennert, K. Wildberger, R.Zeller and P.H.Dederichs,
 Phys. Rev. B{\bf 54} (1996) 14121.
 \bibitem{6Cabria1} I. Cabria, B. Nonas, R. Zeller, and P.H. Dederichs,
 Phys. Rev. B{\bf 65} (2002) 054414.
 \bibitem{6Hjortstam1} O. Hjortstam, J. Trygg, J.M. Wills, B. Johnansson, and O. Eriksson,
 Phys. Rev. B{\bf 53} (1996) 9204.
 \bibitem{6Hjortstam2} O. Hjortstam, K.Baberschke, J.M. Wills, B. Johnansson, and O. Eriksson,
 Phys. Rev. B{\bf 55} (1997) 15026.
 \bibitem{6Trygg1} J. Trygg, J.M. Wills, B. Johnansson, and O. Eriksson,
 Phys. Rev. Lett. {\bf 75} (1995) 2871.
 \bibitem{6Gong1} X.G.Gong and Vijay Kumar,
  Phys. Rev. B{\bf 50} (1994) 17701.
 \bibitem{6Jamneala1} T. Jamneala, V. Madhavan, W. Chen and M. F. Crommie,
  Phys. Rev. B{\bf 61} (2000) 9990.
 \bibitem{Hohenberg1}P.Hohenberg and W.Kohn,
   Phys. Rev. {\bf 136} (1964) B864.
 \bibitem{Kohn1}W.Kohn and L.J.Sham,
 Phys. Rev. {\bf 140} (1965) A1133.
 \bibitem{Blochl1}P.E.Bl\"ochl,
  Phys. Rev. B{\bf 50} (1994) 17953.
 \bibitem{Perdew1} J. P. Perdew, K. Burke,and M. Ernzerhof, Phys. Rev. Lett. {\bf 77} (1996) 3865.
 \bibitem{Kresse1} G.Kresse and J.Furthmuller,
 Comput. Mater. Sci. {\bf 6} (1996) 15.
 \bibitem{Kresse2}G.Kresse and J.Furthmuller,
 Phys. Rev. B{\bf 54} (1996) 11169.
 \bibitem{Kresse3}G.Kresse and D. Joubert,
 Phys. Rev. B{\bf 59} (1999) 1758.
 \bibitem{Hobbs1}D.Hobbs, G.Kesse, J. Hafner,
 Phys. Rev. B{\bf 62} (1994) 11556.
 \bibitem{6Brooks1}M.S.S.Brooks and P.J.Kelly,
 Phys. Rev. Lett. {\bf 51} (1983) 1708.
 \bibitem{6Solovyev1}I.V.Solovyev, A.I.Liechtenstein,K.Terakura,
 Phys. Rev. Lett. {\bf 80} (1998) 5758.
 \bibitem{6Eriksson1}Olle Eriksson, M.S.S.Brooks and B\"orje Johansson,
 Phys. Rev. B. {\bf 41} (1990) 7311.
 \bibitem{6Nonas1}B.Nonas, I. Cabria, R. Zeller, and P.H.Dederichs, T. Huhne and H. Ebert,
 Phys. Rev. Lett. {\bf 86} 2146 (2001) 2146.
 \bibitem{2Daw2} S.W.Daw and M.I.Baskes, Phys.  Rev. B{\bf 29}, (1984) 6443.
 \bibitem{2Ercolessi1} F.Ercolessi, J.B.Adams, EuroPhys. Lett. {\bf 26} (1994) 583.
 \bibitem{6Fan1} The work is parallel to our another work in which we have calculated the magnetism
  of one-dimension Ni chains where the noncollinear magnetsm is important. This is why we have introdced
  the noncollinear calculation in this work. W. Fan and X. G. Gong, (unpublished).
 \bibitem{6Kondo1} J.Kondo, Prog. Theor. Phys. {\bf 32} (1964) 37.
 \bibitem{6Hewson1} A.C. Hewson,
  {\bf The Kondo Problem to Heavy Fermions} Cambridge University Press (1993).
 \bibitem{6Wilson1} K.G.Wilson,
  Rev. Mod. Phys. {\bf 47} (1975) 773.
\end{thebibliography}
\end{document}